\documentclass[twocolumn,superscriptaddress,pre,10pt]{revtex4-2}

\usepackage{amsmath,amssymb}
\usepackage{graphicx}
\usepackage{physics}
\usepackage{bm}
\usepackage[T1]{fontenc}
\usepackage[colorlinks=true, linkcolor=blue, citecolor=blue, urlcolor=blue]{hyperref}
\usepackage{xcolor}

\begin{document}

\title{Mean-field theory of myopic self-avoiding fractional Brownian motion}

\author{Rashad Bakhshizada}
\affiliation{Department of Physics, Missouri University of Science and Technology, Rolla, MO 65409, USA}

\author{Skirmantas Janu\v{s}onis}
\affiliation{Department of Psychological and Brain Sciences, University of California, Santa Barbara, CA 93106, USA}

\author{Ralf Metzler}
\affiliation{Institute of Physics and Astronomy, University of Potsdam, D-14476 Potsdam-Golm, Germany}

\author{Thomas Vojta}
\affiliation{Department of Physics, Missouri University of Science and Technology, Rolla, MO 65409, USA}

\date{\today}

\begin{abstract}
Myopic self-avoiding fractional Brownian motion (FBM) is a stochastic
process in which an ensemble of particles is driven by fractional Gaussian noise while being repelled by the gradient of the
time-integrated ensemble density [J. House et al., Phys.\ Rev.\ E \textbf{112},
034119 (2025)]. Depending on the
anomalous diffusion exponent $\alpha$ characterizing the noise, the
process features two dynamical regimes: an interaction-dominated regime
($\alpha < \alpha_c=4/(d+2)$) where the mean-density interaction governs
long-time dynamics, and a noise-dominated regime ($\alpha > \alpha_c$)
where FBM correlations prevail. In the interaction-dominated regime, the
mean-squared displacement grows as $\langle r^2(t) \rangle \sim t^{4/(d+2)}$
regardless of $\alpha$, while for $\alpha > \alpha_c$ the standard FBM
scaling $\langle r^2(t) \rangle \sim t^{\alpha}$ is recovered. Here, we
develop an analytical mean-field theory of myopic self-avoiding FBM,
based on a Fokker-Planck approach to the interaction-dominated regime.
This allows us to derive closed-form polynomial solutions for the
probability density. To compare with computer simulations, we develop
an efficient radial binning algorithm that significantly  reduces the computational
complexity, making
large-scale three-dimensional simulations feasible. Extensive simulations
in one, two, and three dimensions confirm the analytical predictions. We also discuss the application of the process to the
self-organization of serotonergic axons (fibers) in
vertebrate brains, where FBM paths with self-avoidance
provide a natural framework for understanding spatial
heterogeneities of fiber densities.
\end{abstract}

\maketitle

\section{Introduction}
\label{sec:intro}
Fractional Brownian motion (FBM) is a self-similar, non-Markovian Gaussian
stochastic process characterized by stationary increments with
long-range temporal correlations and power-law scaling of the
mean-squared displacement (MSD) $\langle r^2(t) \rangle \sim t^{\alpha}$ ~\cite{Kolmogorov1940, Mandelbrot1968}\footnote{Throughout this paper, $A \sim B$ denotes asymptotic scaling, i.e., $\lim_{t\to\infty} A/B = \mathrm{const}$.}. It
has emerged as a powerful framework for modeling anomalous
diffusion in diverse systems.
FBM has been applied to anomalous diffusion inside biological
cells~\cite{Szymanski2009, Magdziarz2009, Weber2010, Jeon2011,
Jeon2012, Tabei2013}, the dynamics of
polymers~\cite{Chakravarti1997, Panja2010}, electronic network
traffic~\cite{Mikosch2002}, fluctuations of financial
markets~\cite{Comte1998, Rostek2013}, and the motion of
animals~\cite{Vilk2022}.
Recently, FBM has yielded important insights into the
self-organizing dynamics of serotonergic axons (fibers)
in the brains of all vertebrate animals~\cite{Janusonis2019,
Janusonis2020, Janusonis2023, Janusonis2025}. These fibers
form massive meshworks in species ranging from ancient fishes
(e.g., sharks) to mammals (e.g., humans) and appear to
fundamentally support neuroplasticity~\cite{Miller2026}. Importantly, serotonergic axons
possess considerable physical rigidity~\cite{Bernal2007, deRooij2018} and
grow with pronounced directional persistence, so that successive increments
of a fiber trajectory are correlated rather than independent. This
persistence is captured by the long-range-correlated motion of FBM, whereas
it is absent in an uncorrelated (normal) random walk. Consistent with this
picture, serotonergic fibers tend to accumulate near brain
boundaries~\cite{Janusonis2019, Janusonis2020, Janusonis2023, Janusonis2025},
an accumulation naturally reproduced by repulsive, superdiffusive FBM.
 Reflected FBM can thus bridge the individual
(microscale) fiber trajectories and the emergence of
large-scale fiber-density heterogeneities in neural
tissue~\cite{Janusonis2023, Janusonis2025}. However, FBM
paths are ``blind'' to each other whereas biological axons are
not~\cite{Chen2017, Katori2017}. In particular, axons may
experience a repulsive force in regions that have already
accumulated many fibers. Experimental evidence suggests
that this information may be conveyed to serotonergic fibers
by the local concentration of extracellular serotonin, a
neurotransmitter that the fibers themselves release along
their paths~\cite{Gianni2023, Nazzi2024, Vicenzi2021}.
Motivated by this neurobiological system, as well as by
similar ecological systems with stigmergy~\cite{Theraulaz1999},
we have recently introduced
\textbf{mean-density interactions}: In an ensemble of particles
undergoing FBM, each particle responds to the gradient of the
time-integrated density of the entire ensemble~\cite{House2025}. The resulting process can be understood as a generalization
of the true or myopic self-avoiding random walk~\cite{Amit1983,
Pietronero1983, Peliti1987, Lawler1991}, where trajectories
are repelled from highly visited regions. The process can therefore be called myopic self-avoiding FBM.\par A phenomenological scaling theory of myopic self-avoiding FBM demonstrated a critical value $\alpha_{c}$ of the
exponent characterizing the fractional noise. It separates an  interaction-dominated regime
($\alpha < \alpha_{c}$) from a noise-dominated regime ($\alpha > \alpha_{c}$). This behavior was confirmed by extensive computational simulations in one space dimension~\cite{House2025}.

Here, we develop a mean-field theory of myopic self-avoiding FBM via a Fokker-Planck analysis,
allowing us to obtain analytical solutions for probability densities
in one, two, and three dimensions. In order to compare our analytical
solutions to computational simulations in two and three dimensions, we develop an efficient radial binning algorithm, significantly reducing the computational complexity. We achieve excellent quantitative agreement between theory and simulation across all dimensions.

The paper is organized as follows. Sec.~\ref{sec:model} defines myopic self-avoiding FBM in arbitrary dimensions. Sec.~\ref{sec:scaling}  reviews the scaling theory developed
in Ref.~\cite{House2025} and derives the threshold exponents $\alpha_c$. Sec.~\ref{sec:fokker_planck}
presents the Fokker-Planck formulation. Sec.~\ref{sec:numerics}  details the numerical
methods, including radial binning. In Sec.~\ref{sec:msd}, we demonstrate that the MSD follows the scaling predictions in all dimension. Sec.~\ref{sec:results} presents analytical
solutions and simulation results for one, two, and three dimensions.
We conclude in Sec.~\ref{sec:conclusions} with implications for biological systems
and future extensions.

\section{Myopic self-avoiding FBM}
\label{sec:model}
\subsection{FBM}
To define myopic self-avoiding FBM, we follow Ref.~\cite{House2025}.
FBM is a continuous-time, centered Gaussian
stochastic process for the position $\mathbf{X}(t)$ of a particle
starting at the origin at time $t = 0$~\cite{Kolmogorov1940,
Mandelbrot1968}. In $d$ dimensions, the position has components
$X_\nu(t)$, $\nu = 1, \ldots, d$. The $d$ components are independent
of each other, each following an FBM process with the same anomalous
diffusion exponent $\alpha$. The covariance function of the position
components $X_\nu$ at times $s$ and $t$ is given by
\begin{equation}
\langle X_\nu(s)\, X_{\mu}(t) \rangle =
\left(s^\alpha - |s-t|^\alpha + t^\alpha\right)K\,\delta_{\nu\mu},
\label{eq:fbm_covariance_cont}
\end{equation}
with the anomalous diffusion exponent $\alpha \in (0,2]$ and
the generalized diffusion coefficient $K$.
$\delta_{\nu\mu}$ denotes the Kronecker delta ($\delta_{\nu\mu} = 1$ if $\mu = \nu$ and
$\delta_{\nu\mu} = 0$ otherwise). The anomalous diffusion exponent
is related to the alternatively used Hurst exponent $H$ via $\alpha = 2H$. Setting $s=t$ gives
$\langle X^2(t) \rangle = 2dKt^\alpha$, confirming anomalous
diffusion. For $\alpha = 1$, FBM reduces to normal Brownian motion;
$\alpha > 1$ corresponds to superdiffusion with persistent increments,
while $\alpha < 1$ gives subdiffusion with anti-persistent increments.

For computer simulations, we work with a discrete-time version of
FBM~\cite{Qian2003}. We discretize time by setting
$\mathbf{r}_n = \mathbf{X}(t_n)$ with $t_n = \epsilon n$,
where $\epsilon$ is the time step and $n$ is a non-negative integer.
The time evolution takes the form of a discrete-time random walk with correlated steps,
\begin{equation}
\mathbf{r}_n = \mathbf{r}_{n-1} + \bm{\xi}_n,
\label{eq:recursion_single}
\end{equation}
where $\bm{\xi}_n$ is a $d$-dimensional fractional Gaussian noise
with zero mean, step variance $\sigma^2 = 2K\epsilon^\alpha$, and
covariance
\begin{equation}
\begin{split}
\langle \xi_{m,\nu}\, \xi_{m+n,\mu} \rangle &= C_{n} \,\delta_{\nu\mu} \\
&= \frac{1}{2}\sigma^2\left(|n+1|^\alpha - 2|n|^\alpha
+ |n-1|^\alpha\right)\delta_{\nu\mu}~.~
\end{split}
\label{eq:covariance}
\end{equation}
For $n \to \infty$, the covariance decays as
$C_n \sim \alpha(\alpha-1)|n|^{\alpha-2}$, which is positive
(persistent) for $\alpha > 1$ and negative (anti-persistent)
for $\alpha < 1$. For $\alpha = 1$, $C_n = 0$ for all $n \neq 0$,
recovering uncorrelated Brownian motion.

\subsection{Mean-Density Interaction}

We now consider an ensemble of $N$ particles, each undergoing an
independent FBM process, but subject to a force that couples each
particle to the gradient of the time-integrated mean density:
\begin{equation}
\mathbf{r}_{n}^{(j)} = \mathbf{r}_{n-1}^{(j)} + \bm{\xi}_{n}^{(j)}
+ \mathbf{f}(\mathbf{r}_{n-1}^{(j)}, t_{n-1}),
\label{eq:recursion}
\end{equation}
where the force term is given by
\begin{equation}
\mathbf{f}(\mathbf{r}_n^{(j)}, t_n) = -A \nabla
P_{\text{int}}(\mathbf{r},t_n)\bigg|_{\mathbf{r}=\mathbf{r}_n^{(j)}}.
\label{eq:force}
\end{equation}
Here, $j = 1, \ldots, N$ labels the particles in the ensemble,
and the mean time-integrated density is defined as
\begin{equation}
P_{\rm int}(\mathbf{r},t_n) = \frac{\epsilon}{N}\sum_{j=1}^{N}\sum_{m=1}^{n}
\delta\big(\mathbf{r}-\mathbf{r}^{(j)}_m\big).
\label{eq:Pint}
\end{equation}
Here, $\delta(\mathbf{r} - \mathbf{r}_m^{(j)})$ is the
$d$-dimensional Dirac delta function.
The normalization of $P_{\text{int}}$ grows linearly with  discrete time,
\begin{equation}
\int P_{\rm int}(\mathbf{r},t_n)\, d^d r = \epsilon\, n = t_n,
\label{eq:Pint_norm}
\end{equation}
reflecting the accumulation of trajectories over time. The positive
coupling constant $A$ in the mean-density force causes particles to avoid regions of high density \footnote{The force in Eq.~(\ref{eq:force}) depends on the density
accumulated up to the current time. FBM (like Brownian motion) is invariant
under time reversal in distribution: for any fixed time $T$, the reversed
path $X(T-t)-X(T)$, $0\le t\le T$, has the same probability law as $X(t)$.
The process defined by Eqs.~(\ref{eq:recursion})--(\ref{eq:Pint}) does not
have this symmetry, because the mean-density interaction couples each step
to the accumulated past of the ensemble.}.
\subsection{Mean-Field Limit}

For finite ensemble size $N$, the integrated density
\eqref{eq:Pint} is expected to fluctuate from
ensemble to ensemble. In the present paper, we will
focus on the infinite-ensemble-size (mean-field) limit,
$N \to \infty$. In this limit, the ensemble-to-ensemble
fluctuations of $P_{\text{int}}$ will be suppressed. Consequently,
$P_{\mathrm{int}}$ of an individual ensemble becomes
identical to the ensemble average. This will allow us
to develop an analytical approach in Sec.~\ref{sec:fokker_planck}.
Finite-$N$ effects were studied numerically in $d = 1$
in Ref.~\cite{House2025}.

In two and three dimensions, the time evolution of myopic self-avoiding
FBM preserves rotational symmetry \textit{in the statistical sense}.
That is, if the initial conditions are rotationally invariant --- as is
fulfilled when all particles start at the origin --- then the statistical
properties of the process remain rotationally invariant at all later
times. As a consequence, we expect $P_{\text{int}}$ to depend only on
$r = |\mathbf{r}|$ in the mean-field limit, and not on the direction of $\mathbf{r}$. In two dimensions ($d=2$), this yields polar symmetry, and in three dimensions ($d=3$), spherical symmetry emerges. These symmetries simplify both the analytical treatment
and the numerical simulations, which we will exploit in subsequent sections.

\section{Scaling theory}
\label{sec:scaling}
In this section, we summarize the self-consistent phenomenological scaling theory
developed in Ref.~\cite{House2025}. For sufficiently long times,
we assume that the integrated density in $d$ dimensions fulfills
the scaling form
\begin{equation}
P_{\text{int}}(\mathbf{r},t) = \frac{t}{[b(t)]^d} Y\left(\frac{|\mathbf{r}|}{b(t)}\right).
\label{eq:scaling_Pint}
\end{equation}
This scaling form expresses the expectation that, at long times, the
integrated density does not change its shape: as time progresses, the
profile of $P_{\mathrm{int}}$ is merely stretched by a characteristic length
scale $b(t)\sim t^{\delta}$, with $\delta$ the (so far) unknown scaling
exponent, while its amplitude grows such that the normalization is
preserved \footnote{The scaling form~\eqref{eq:scaling_Pint} does not
contain $\alpha$: it encodes scaling with a single length $b(t)$ and the
normalization~\eqref{eq:Pint_norm}. It thus applies to normal Brownian motion
as well as FBM.}. The prefactor $t/[b(t)]^d$ in Eq.~\eqref{eq:scaling_Pint}
is fixed by the normalization: in the continuum limit, the discrete-time
condition~\eqref{eq:Pint_norm} becomes $\int P_{\mathrm{int}}\,d^d r = t$.
Inserting the scaling form~\eqref{eq:scaling_Pint} gives
$\int P_{\mathrm{int}}\,d^d r = t\int Y(y)\,d^d y$, which reproduces this
linear growth in time when the scaling function is normalized to
\begin{equation}
\int_0^\infty Y(y)\, d^dy=
\Omega_d \int_0^\infty y^{d-1} Y(y)\,dy = 1,
\label{eq:Y_norm}
\end{equation}

with $\Omega_d = 2\pi^{d/2}/\Gamma(d/2)$ the surface area of the
unit sphere in $d$ dimensions. As the scaling form of
$P_{\text{int}}$ is rotationally invariant, the resulting forces
have a radial component only. The radial force component scales as
\begin{equation}
f_r(r,t) \sim -\frac{At}{[b(t)]^{d+1}} Y'\left(\frac{r}{b(t)}\right)
\sim t^{1-(d+1)\delta},
\end{equation}
where $Y'$ denotes the derivative of the scaling function $Y$.
If the interactions dominate the motion, the typical displacement
is given by the integral over the force,
\begin{equation}
r_{\text{typ}}(t) \sim \int_0^t \tau^{1-(d+1)\delta}\,d\tau
\sim t^{2-(d+1)\delta}.
\end{equation}
Self-consistency ($r_{\text{typ}} \sim b(t) \sim t^\delta$) yields
\begin{equation}
\delta = \frac{2}{d+2}.
\label{eq:scaling_exponents}
\end{equation}
The MSD is expected to scale as $b^2(t)$, yielding
\footnote{We distinguish the exponent $\alpha$ that
parametrizes the fractional {Gaussian} noise from the exponent
$\bar{\alpha}$ that characterizes the asymptotic power-law growth
of the MSD, $\langle r^2(t)\rangle \sim
t^{\bar{\alpha}}$, in the interaction-dominated regime.}
\begin{equation}
\langle r^2(t) \rangle \sim t^{\bar{\alpha}}, \qquad
\bar{\alpha} = 2\delta = \frac{4}{d+2}.
\end{equation}
The phenomenological scaling theory thus suggests the following
picture. The threshold exponent $\alpha_c = 4/(d+2)$ separates two
distinct dynamical regimes. For $\alpha < \alpha_c$, the mean-density
interaction dominates the long-time dynamics, imposing the universal
MSD exponent $\bar{\alpha} = \alpha_c$ regardless of the value of
$\alpha$. Consequently, $\langle r^2(t) \rangle \sim t^{4/(d+2)}$,
independent of $\alpha$. For $\alpha > \alpha_c$, the FBM noise
dominates and $\bar{\alpha} = \alpha$, recovering standard FBM
scaling $\langle r^2(t) \rangle \sim t^\alpha$. For specific
dimensions: $d=1$ gives $\delta = 2/3$, $\alpha_c = 4/3$;
$d=2$ gives $\delta = 1/2$, $\alpha_c = 1$; and $d=3$ gives
$\delta = 2/5$, $\alpha_c = 4/5$.

\section{Fokker-Planck approach}
\label{sec:fokker_planck}

For the special case of $\alpha = 1$, myopic self-avoiding FBM
reduces to myopic self-avoiding (normal) Brownian motion. In this
case, the process can be described by a Fokker-Planck equation for
the instantaneous probability density $P(\mathbf{r},t)$. To derive
it, we start from the continuity equation
\begin{equation}
    \frac{\partial P}{\partial t} + \nabla \cdot \mathbf{J} = 0,
\end{equation}
where the probability current has a drift contribution from the
interaction force $\mathbf{F} = -A\nabla P_{\mathrm{int}}$ and a
diffusive contribution from the noise,
\begin{equation}
    \mathbf{J} = P\mathbf{F} - D\nabla P
               = -AP\,\nabla P_{\mathrm{int}} - D\nabla P.
\end{equation}
Substituting into the continuity equation gives the Fokker-Planck
equation
\begin{equation}
    \frac{\partial P}{\partial t}
    = A\nabla \cdot [P \nabla P_{\text{int}}] + D\nabla^2 P.
    \label{eq:fokker_planck}
\end{equation}
The relation between $P_{\mathrm{int}}$ and $P$ follows from the
continuum limit of Eq.~\eqref{eq:Pint},
\begin{equation}
    P_{\mathrm{int}}(\mathbf{r},t) = \int_0^t P(\mathbf{r},\tau)\,d\tau,
\end{equation}
or, equivalently,
\begin{equation}
\frac{\partial P_{\mathrm{int}}}{\partial t} = P.
\label{eq:consistency_main}
\end{equation}
In the long-time regime, the solutions of the Fokker-Planck
equation~\eqref{eq:fokker_planck} are expected to fulfill the
scaling theory of Sec.~\ref{sec:scaling}. This means $P_{\mathrm{int}}$ is expected to
fulfill Eq.~\eqref{eq:scaling_Pint}. The corresponding scaling
form for $P$ reads
\begin{equation}
P(r,t) = [b(t)]^{-d}\, Z(r/b(t))
\label{eq:scaling_P}
\end{equation}
with $b(t) \sim t^\delta$. The consistency relation~\eqref{eq:consistency_main} between the instantaneous density
and the integrated density then implies
\begin{equation}
    Z(y) = (1-d\delta)Y(y) - \delta y Y'(y).
    \label{eq:consistency}
\end{equation}
This relation will be used in Sec.~\ref{sec:results} to derive
analytical solutions  in each dimension.

For general FBM with $\alpha \neq 1$, no local Fokker-Planck
equation exists due to the non-Markovian nature of fractional
Gaussian noise. Note that Lutz~\cite{Lutz2001} presented a diffusion equation
that yields the correct Green's function for FBM in the absence
of boundary conditions. However, it is actually the diffusion
equation for scaled Brownian motion, a fundamentally different
stochastic process (see Ref.~\cite{Jeon2014} for a detailed discussion), and solving it with nontrivial boundary conditions
does not give the correct results for FBM. Recent progress
in the formulation of a generalized diffusion equation
for non-Markovian processes is reported in
Ref.~\cite{Taloni2025}.

For myopic self-avoiding FBM, in the interaction-dominated
regime ($\alpha < \alpha_c$), the deterministic forces dominate
over the noise at long times. In this regime, an ``interaction Fokker-Planck equation'' without the diffusion term,
\begin{equation}
    \frac{\partial P}{\partial t}
    = A\nabla \cdot [P \nabla P_{\text{int}}]~,
    \label{eq:fokker_planck_2}
\end{equation}
provides an appropriate
description of the long-time behavior (and the consistency relation~\eqref{eq:consistency} still holds).

\section{Numerical simulations}
\label{sec:numerics}

We perform simulations of myopic self-avoiding FBM in one, two, and
three dimensions. We set the time step $\epsilon = 1$ and the
generalized diffusion coefficient $K = 1/2$, which gives a step
variance $\sigma^2 = 2K\epsilon^\alpha = 1$ for all $\alpha$.
Ensemble sizes range from $N = 8{,}192$ to $N = 65{,}536$
particles, and the probability densities are averaged over
$64$ to $256$ independent ensembles. Simulations are carried
out up to a maximum time of $N_{T}= 2^{22}$ time steps. We have explored
a range of interaction strengths from $A=0.25$ to 4. In the  simulations presented here,
the interaction strength takes the values $A = 0.25$ in one dimension,
$A = 1$ in two dimensions, and $A = 0.25$ in three dimensions. Other values give analogous results.

Fractional Gaussian noise is generated via the Fourier-filtering
method~\cite{Makse1996}. This method starts from a sequence of
independent Gaussian random numbers, $\chi_{n}$, computes their Fourier
transform $\tilde{\chi}_\omega$, and multiplies it by
$[\tilde{C}(\omega)]^{1/2}$, where $\tilde{C}(\omega)$ is the
Fourier transform of the covariance function~\eqref{eq:covariance}.
The inverse Fourier transform then yields the desired correlated
noise sequence $\xi_{n}$ with  the correct long-range statistics.

At each time step, the integrated density $P_{\mathrm{int}}$ is
updated using a histogram with cell size $\Delta r$ ranging
from $0.1$ to $1.0$ depending on the dimensionality and
system size. To avoid
the singularity in the delta function appearing in the definition
of $P_{\mathrm{int}}$, Eq.~\eqref{eq:Pint}, each particle's
contribution is broadened by a Gaussian of width
$\sigma_w = 0.5$. Forces are then computed from the gradient
of $P_{\mathrm{int}}$ using fourth-order finite differences, and
particle positions are advanced according to
Eq.~\eqref{eq:recursion}. In Ref.~\cite{House2025}, the dependence
of the simulation results on the values of $\Delta r$ and $\sigma_w$
was tested. Only minor changes were observed.

In two and three dimensions, simulations are limited by the
computer memory required to store the two-dimensional or
three-dimensional density array with sufficient resolution.
To overcome this limitation, we exploit the rotational symmetry
of the density and introduce a radial binning algorithm,
detailed in Appendix~\ref{sec:appendix}. Instead of storing the
full $d$-dimensional density array, we bin the density as a
function of the radial coordinate $r = |\mathbf{r}|$ only,
reducing the required computer memory from
$\mathcal{O}(L^d)$ for a Cartesian grid to $\mathcal{O}(L)$. Here $L$ is the spatial system size.
This makes large-scale three-dimensional simulations feasible.
The validity of the radial binning approach is confirmed by
direct comparison with Cartesian binning in two dimensions in the Appendix~\ref{sec:appendix} .

During each simulation, we measure three quantities, the MSD $\langle r^2(t) \rangle$, the time-integrated density
$P_{\mathrm{int}}(r,t)$, and the instantaneous probability density $P(r,t)$. To reduce the statistical errors, that stem from
ensemble-to-ensemble fluctuations due to the finite
ensemble size,  $P$ is measured by time-averaging over the
last $4\%$ of the simulation (from $t = 0.96\,N_T$ to
$t = N_T$), after the system has reached its stationary
scaling regime.


\section{Mean-squared displacement}
\label{sec:msd}

The phenomenological scaling theory of Ref.~\cite{House2025}
predicts two dynamical regimes separated by the threshold exponent
$\alpha_c = 4/(d+2)$: an interaction-dominated regime for
$\alpha < \alpha_c$ where $\langle r^2(t) \rangle \sim t^{4/(d+2)}$
regardless of $\alpha$, and a noise-dominated regime for
$\alpha > \alpha_c$ in which the standard FBM behavior
$\langle r^2 \rangle \sim t^\alpha$ is recovered. In one
dimension, this picture was confirmed by computer simulations
\cite{House2025}: For $\alpha < \alpha_c = 4/3$, the
MSD was found to grow as
$\langle x^2(t) \rangle \sim t^{4/3}$, independent of $\alpha$,
while for $\alpha > \alpha_c$ the FBM scaling
$\langle x^2(t) \rangle \sim t^\alpha$ was observed.

\begin{figure}
    \centering
    \includegraphics[width=\linewidth]{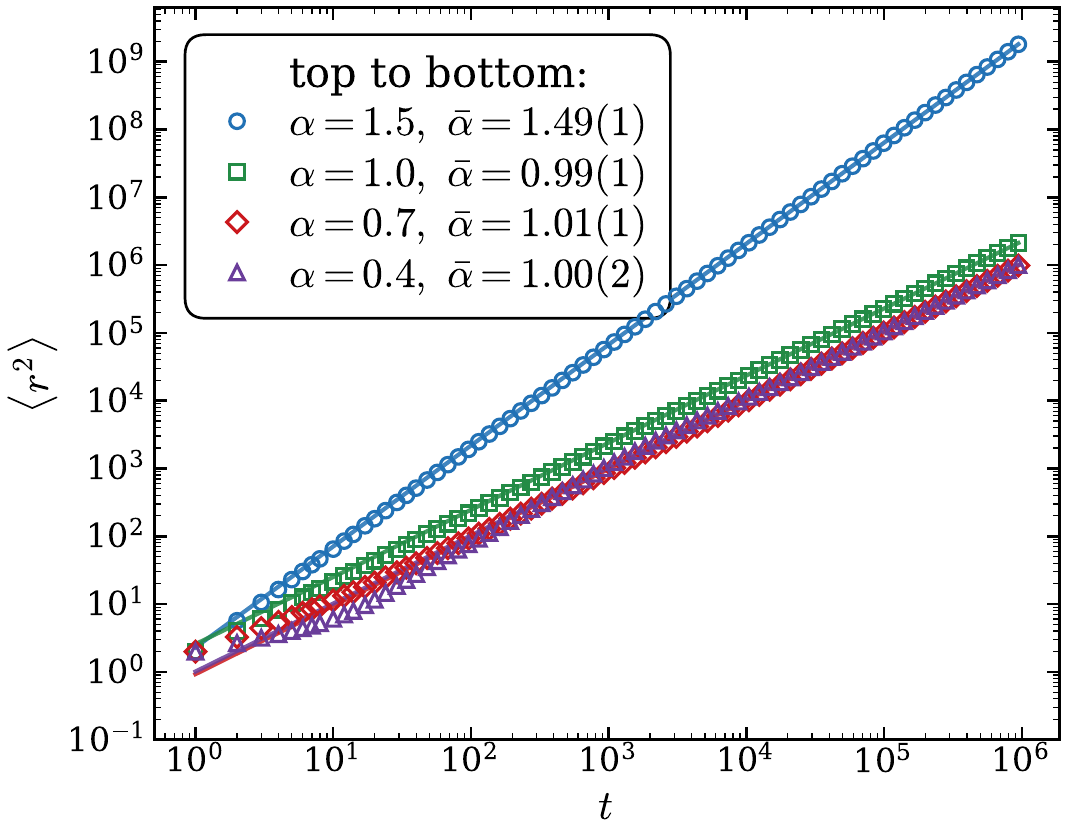}
    \caption{MSD $\langle r^2 \rangle$
    vs time $t$ for myopic self-avoiding FBM in two dimensions,
    for four values of $\alpha$. The statistical errors of the data points are
    smaller than the symbol size. For $\alpha = 1.5 > \alpha_c = 1$,
    the MSD grows as $\langle r^2 \rangle \sim t^{\alpha}$,
    recovering standard FBM scaling. For $\alpha = 1.0$, $0.7$,
    and $0.4$, all at or below $\alpha_c = 1$, the asymptotic MSD exponent
    $\bar{\alpha}$ equals $\alpha_c$, regardless of
    $\alpha$, demonstrating that the mean-density interaction
    dominates the long-time dynamics. The values of $\bar{\alpha}$ are
    extracted from weighted power-law fits to the long-time behavior
    (solid lines). Their error represents the sum  of statistical errors
    and systematic errors from choosing the fit range, and are given
    in units of the last digit, e.g., $\bar{\alpha} = 0.99(1)$
    means $\bar{\alpha} = 0.99 \pm 0.01$.}
    \label{fig:msd_2D}
\end{figure}

Here, we extend this analysis to two and three dimensions
using our large-scale simulations. Figure~\ref{fig:msd_2D}
shows $\langle r^2(t) \rangle$ as a function of $t$ in two dimensions
for four values of $\alpha$.
For $\alpha = 1.5 > \alpha_c = 1$,
the MSD grows as $\langle r^2(t) \rangle \sim t^{1.5}$, confirming
that the noise-dominated regime recovers standard FBM scaling.
For $\alpha = 1.0$, $0.7$, and $0.4$, all at or below
$\alpha_c = 1$, the MSD follows the power law
$\langle r^2 \rangle \sim t^{\bar{\alpha}}$, with the MSD exponent $\bar{\alpha}$
asymptotically  behaving as $\bar{\alpha} = 1 = \alpha_c$=$2\delta$, independent
of $\alpha$ \cite{Note4}.
The two sub-critical curves collapse onto each other for long
$t$, demonstrating that the mean-density interaction dominates
the long-time dynamics, whereas the noise is subleading. In the marginal case $\alpha = \alpha_c = 1$,
$\langle r^2 \rangle \sim t$, i.e., $\bar{\alpha} = 1$, but
with a larger prefactor, as both noise and interactions
contribute to the motion.

\begin{figure}
    \centering
    \includegraphics[width=\linewidth]{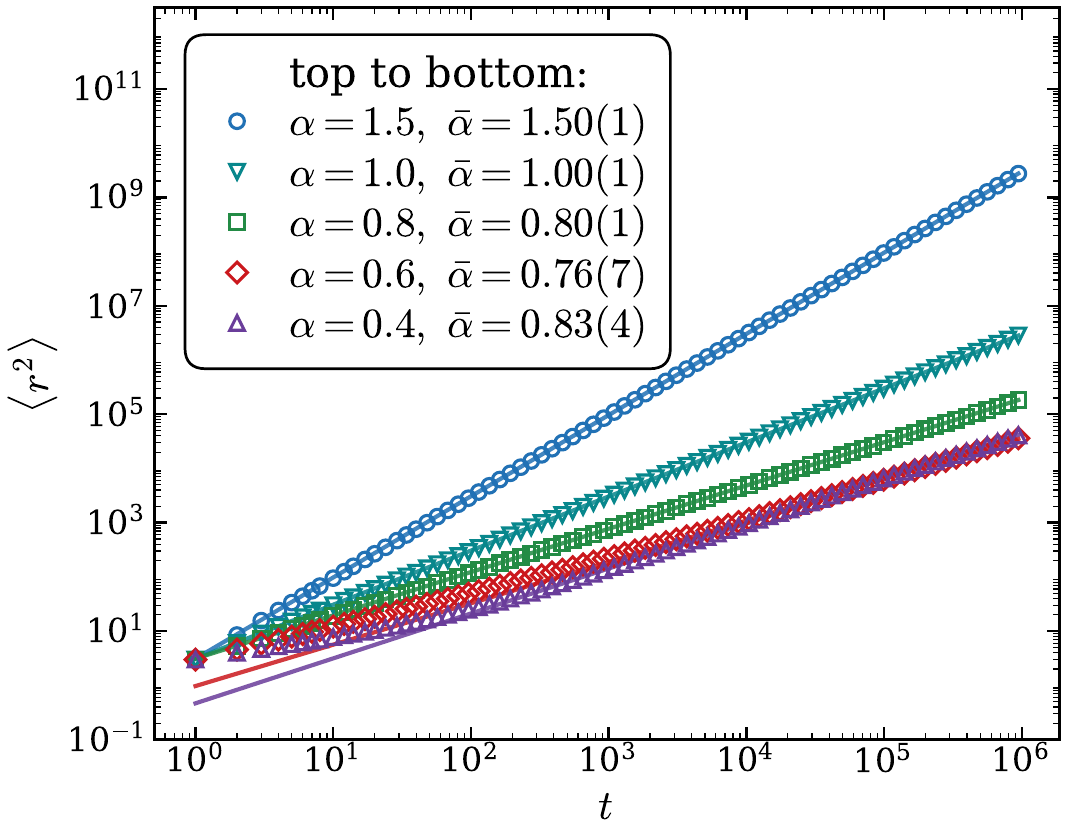}
    \caption{MSD $\langle r^2 \rangle$
    vs time $t$ for myopic self-avoiding FBM in three dimensions,
    for five values of $\alpha$. For $\alpha = 1.5$ and 1.0, both above $\alpha_c = 4/5$,
    the MSD recovers the standard FBM scaling
    $\langle r^2 \rangle \sim t^\alpha$. For $\alpha = 0.8$, $0.6$,
    and $0.4$, all at or  below $\alpha_c = 4/5$, the fitted exponent
    $\bar{\alpha}$ saturates at $\bar{\alpha} = \alpha_c = 4/5$
    regardless of $\alpha$, confirming the interaction-dominated
    regime and the scaling prediction $\delta = 2/5$. The values of $\bar{\alpha} $
    are extracted from weighted power-law fits to the long-time behavior (solid lines).
    The statistical errors of the data points are
    smaller than the symbol size.}
    \label{fig:msd_3D}
\end{figure}

Figure~\ref{fig:msd_3D} shows analogous results
in three dimensions for five values of $\alpha$.
For $\alpha = 1.5$ and 1.0, both above $\alpha_c = 4/5$, the noise-dominated scaling
$\langle r^2 (t)\rangle \sim t^{\alpha}$ is recovered. For
$\alpha = 0.8$, $0.6$, and $0.4$, all at or  below $\alpha_c = 4/5$,
the MSD exponent saturates at $\bar{\alpha}=\alpha_{c}=2\delta=4/5$,
independent of $\alpha$ and consistent with the theoretical
prediction $\delta = 2/(d+2) = 2/5$. The saturation of the
MSD exponent in both 2D and 3D provides strong confirmation
that the scaling theory of Ref.~\cite{House2025} correctly
describes the interaction-dominated regime across all dimensions.

\section{Analytical and numerical probability densities}

\label{sec:results}

In this section, we derive analytical solutions of the long-time behavior of myopic self-avoiding FBM. We focus on the interaction-dominated regime because, in the noise-dominated regime, the process becomes identical to non-interacting  FBM for $t \to \infty$. Therefore, we look for solutions of the interaction Fokker-Planck
equation~\eqref{eq:fokker_planck_2} that fulfill the scaling
forms~\eqref{eq:scaling_Pint} and~\eqref{eq:scaling_P} for
$P_{\mathrm{int}}$ and $P$, respectively. Inserting these scaling
forms into the Fokker-Planck equation and applying the consistency
relation~\eqref{eq:consistency} leads to a nonlinear second-order ordinary differential equation (ODE) for the
scaling function $Y(y)$.  We solve the
ODE via the quadratic ansatz $Y(y) = a_0 + a_2 y^2$ (the same
solution can also be obtained using integrating factors). The normalization condition
\begin{equation}
\int Y(y)\, d^dy = 1
\end{equation}
is used to fix the coefficients in the solution.
The instantaneous density $Z(y)$ then follows from the
consistency relation~\eqref{eq:consistency}.

\subsection{One Dimension}

With $d=1$, $\delta = 2/3$, and $\alpha_c = 4/3$, the consistency
relation~\eqref{eq:consistency} between $P_{\mathrm{int}}$ and $P$ gives $Z(y) = \tfrac{1}{3}Y(y) -
\tfrac{2}{3}yY'(y)$. Inserting the scaling forms into Eq. \eqref{eq:fokker_planck_2} and applying the
consistency relation yields the nonlinear second-order ODE for the scaling function $Y(y)$,
\begin{equation}
Y - 3yY' - 2y^2Y'' = -3A\left[Y''Y - 4yY'Y'' - (Y')^2\right].
\label{eq:ODE_1d}
\end{equation}
 The quadratic ansatz
$Y(y) = a_0 + a_2 y^2$ exactly satisfies
Eq.~\eqref{eq:ODE_1d} with $a_2 = -1/(3A)$. As $Y(y)$ must be nonnegative, $y$ is restricted to the range $-y_{\max} \leq y \leq y_{\max}$. The normalization condition
$\int_{-\infty}^\infty Y(y)\,dy = 1$ then fixes
$y_{\max} = (9A/4)^{1/3}$. The resulting scaling function of the  integrated
density is given by
\begin{equation}
Y(y) = \begin{cases}
\dfrac{y_{\max}^2 - y^2}{3A} & |y| \leq y_{\max}\\[6pt]
0 & |y| > y_{\max},
\end{cases}
\label{eq:Y_1d}
\end{equation}
i.e., a downward parabola. In terms of the
original coordinates $x$ and $t$, the integrated density reads
\begin{equation}
\begin{split}
P_{\mathrm{int}}(x, t) &= t^{1/3}\, Y\!\left(\frac{x}{t^{2/3}}\right) \\
&= \begin{cases}
\dfrac{t^{1/3}}{3A}\left(y_{\max}^2 - \dfrac{x^2}{t^{4/3}}\right)
& |x| \leq y_{\max}\, t^{2/3}\\[10pt]
0 & |x| > y_{\max}\, t^{2/3}
\end{cases}
\end{split}
\label{eq:Pint_1d}
\end{equation}
which peaks at the origin and decays parabolically to zero at
$|x| = y_{\max}\, t^{2/3}$. This sharp cutoff is the
interaction-imposed boundary of the accessible region, growing
as $t^{2/3}$ in time.

Substituting Eq.~\eqref{eq:Y_1d} into the consistency
relation~\eqref{eq:consistency} gives the scaling function of the instantaneous
density as
\begin{equation}
Z(y) = \begin{cases}
\dfrac{y_{\max}^2 + 3y^2}{9A} & |y| \leq y_{\max}\\[6pt]
0 & |y| > y_{\max}.
\end{cases}
\label{eq:Z_1d}
\end{equation}
In terms of original coordinates, the instantaneous probability
density reads
\begin{equation}
\begin{split}
P(x,t) &= t^{-2/3}\, Z\!\left(\frac{x}{t^{2/3}}\right) \\
&= \begin{cases}
\dfrac{t^{-2/3}}{9A}\left(y_{\max}^2 + \dfrac{3x^2}{t^{4/3}}\right)
& |x| \leq y_{\max}\, t^{2/3}\\[10pt]
0 & |x| > y_{\max}\, t^{2/3}
\end{cases}
\end{split}
\label{eq:P_1d}
\end{equation}
Unlike $P_{\mathrm{int}}$, the instantaneous density
$P(x,t)$ \emph{increases} with $|x|$, reflecting the fact that
particles are repelled from the dense central region and accumulate
near the outer boundary of the accessible region. We note that results
similar to eqs.\ (\ref{eq:Pint_1d})  and (\ref{eq:P_1d}) were recently obtained in other contexts
\cite{Maggs26,Pineauetal25}.

Simulations of myopic self-avoiding FBM in one dimension were already reported in
Ref.~\cite{House2025}, where the MSD
$\langle x^2(t) \rangle \sim t^{4/3}$ was confirmed to agree with
the scaling theory prediction for $\alpha < \alpha_c = 4/3$.
Here, we therefore focus on showing that the probability densities
$P_{\mathrm{int}}$ and $P$ agree with the Fokker-Planck predictions
derived above.

\begin{figure}
    \centering
    \includegraphics[width=\linewidth]{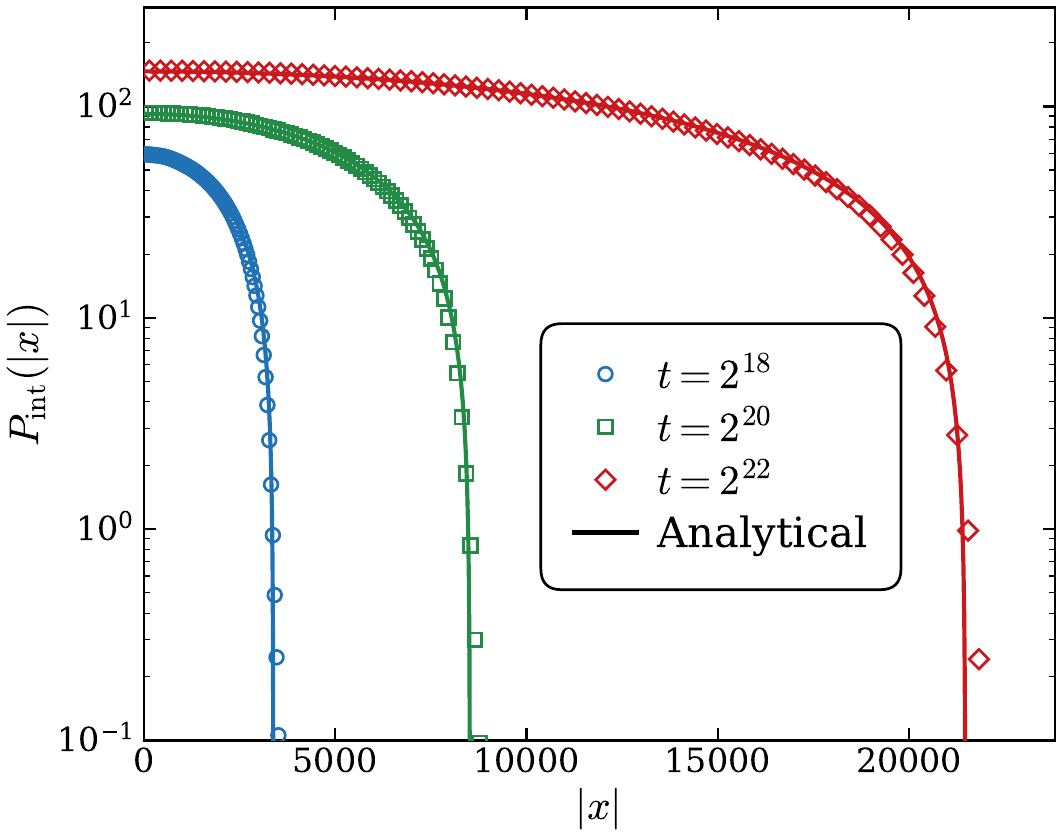}
    \caption{1D integrated probability density $P_{\mathrm{int}}(|x|)$
    vs $|x|$ for $\alpha = 1.0 < \alpha_c = 4/3$. The analytical
    solution Eq.~\eqref{eq:Pint_1d} (solid lines) agrees quantitatively
    with simulations (symbols) at three times $t = 2^{18}$, $2^{20}$,
    $2^{22}$. The density peaks at the origin and decays parabolically
    to zero at $|x| = y_{\max}t^{2/3}$, confirming the predicted
    scaling.}
    \label{fig:1D_integ}
\end{figure}

\begin{figure}
    \centering
    \includegraphics[width=\linewidth]{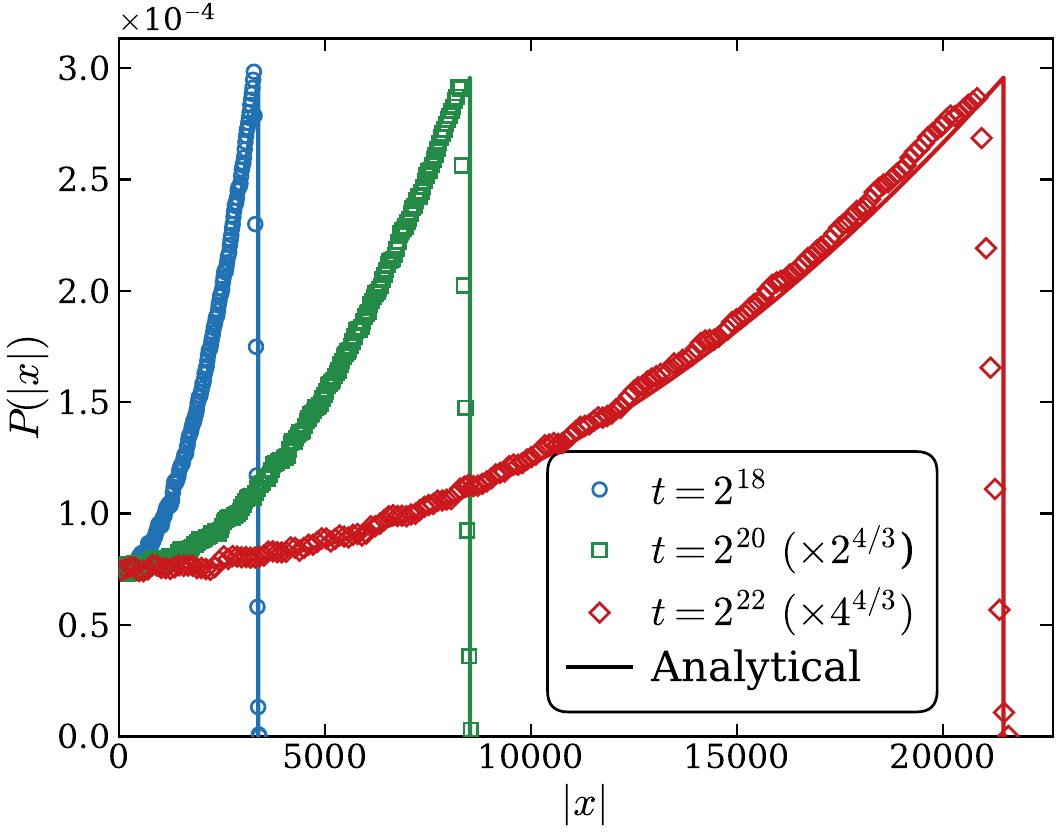}
    \caption{1D instantaneous probability density $P(x,t)$
    vs $|x|$ for $\alpha = 0.7 < \alpha_c = 4/3$, compared to the analytical solution~\eqref{eq:P_1d}. Curves for
    $t = 2^{20}$ and $t = 2^{22}$ are rescaled by the theoretically
    predicted factors $(t/2^{18})^{2/3} \approx 2.52$ and
    6.35, respectively, to improve visibility.
    The data agree well with the characteristic bimodal, U-shaped structure --- with
    particles depleted at the origin and accumulating near the
    outer boundary $|x| = y_{\max} t^{2/3}$. Small differences between theory and
    simulation can be attributed to the effects of finite time and finite ensemble size.}
    \label{fig:1D_inst}
\end{figure}

Figure~\ref{fig:1D_integ} shows the integrated density
$P_{\mathrm{int}}(|x|)$ for $\alpha = 1.0 < \alpha_c = 4/3$,
plotted against the analytical solution Eq.~\eqref{eq:Y_1d}
at three simulation times $t = 2^{18}$, $2^{20}$, and $2^{22}$.
Figure~\ref{fig:1D_inst} shows the
corresponding instantaneous density $P(|x|)$; the curves for
$t = 2^{20}$ and $t = 2^{22}$ are rescaled by the theoretically
predicted factors $(t/2^{18})^{2/3}$, equal to about 2.52
and 6.35, respectively, for better visibility. The excellent agreement between the analytical solutions
Eq.~\eqref{eq:Pint_1d} and Eq.~\eqref{eq:P_1d} and the simulations confirms the
Fokker-Planck predictions for the integrated and  instantaneous densities.

\subsection{Two Dimensions}

With $d=2$, $\delta = 1/2$, $\alpha_c = 1$, the consistency
relation~\eqref{eq:consistency} simplifies to $Z(y) = -yY'/2$,
and the normalization condition is $2\pi \int_0^\infty y Y(y)\,dy = 1$.
The 2D case is special because the threshold exponent $\alpha_c = 1$
coincides with normal Brownian motion. This allows us to study not only the interaction-dominated
regime ($\alpha < \alpha_c$) using the interaction Fokker-Planck equation~\eqref{eq:fokker_planck_2} but also
the marginal case ($\alpha = \alpha_c$), using the Fokker-Planck equation~\eqref{eq:fokker_planck}. In the following, we treat
the two cases, $\alpha < \alpha_c$ (interaction-dominated, noise negligible) and  $\alpha = \alpha_c$ (with diffusion term)
separately.

\subsubsection{Interaction-dominated regime ($\alpha < \alpha_c = 1$)}

For $\alpha < 1$ the diffusion term is negligible and the
interaction Fokker-Planck equation~\eqref{eq:fokker_planck_2}
provides an appropriate description. Inserting the scaling
forms~\eqref{eq:scaling_Pint} and~\eqref{eq:scaling_P} for
$P_{\mathrm{int}}$ and $P$ and applying the consistency
relation~\eqref{eq:consistency}, we arrive at the nonlinear
second-order ODE
\begin{equation}
-\frac{3}{2}y^2 Y' - \frac{1}{2}y^3 Y'' =
2A\,y(Y')^2 + 2A\,y^2 Y' Y'',
\label{eq:ODE_2d_noD}
\end{equation}
for the scaling function $Y(y)$. The quadratic ansatz $Y(y) = a_0 + a_2 y^2$ yields
$a_2 = -1/(4A)$, and $a_0 = y_{\max}^2/4A$, with $y_{\max}$ fixed by the normalization condition
$2\pi\int_0^\infty y Y(y)\,dy = 1$. This yields $y_{\max} = (8A/\pi)^{1/4}$. The solutions are
\begin{align}
Y(y) &= \begin{cases}
\dfrac{y_{\max}^2 - y^2}{4A} & y \leq y_{\max}\\[6pt]
0 & y > y_{\max},
\end{cases}
\label{eq:Y_2d_noD}\\[6pt]
Z(y) &= \frac{y^2}{4A}.
\label{eq:Z_2d_noD}
\end{align}
In terms of original coordinates, the time-integrated density reads
\begin{equation}
\begin{split}
P_{\mathrm{int}}(r,t) &= Y\!\left(\frac{r}{t^{1/2}}\right) \\
&= \begin{cases}
\dfrac{1}{4A}\left(y_{\max}^2 - \dfrac{r^2}{t}\right)
& r \leq y_{\max}\,t^{1/2}\\[10pt]
0 & r > y_{\max}\,t^{1/2},
\end{cases}
\end{split}
\label{eq:Pint_2d}
\end{equation}
and the instantaneous density reads
\begin{equation}
\begin{split}
P(r,t) &= t^{-1}\,Z\!\left(\frac{r}{t^{1/2}}\right) \\
&= \begin{cases}
\dfrac{r^2}{4A\,t^{2}}
& r \leq y_{\max}\,t^{1/2}\\[10pt]
0 & r > y_{\max}\,t^{1/2}.
\end{cases}
\end{split}
\label{eq:P_2d}
\end{equation}
which vanishes at the origin and grows outward, forming an annular
(ring-shaped) particle distribution with a sharp cutoff at
$r = y_{\max}\,t^{1/2}$.

\begin{figure}
    \centering
    \includegraphics[width=\linewidth]{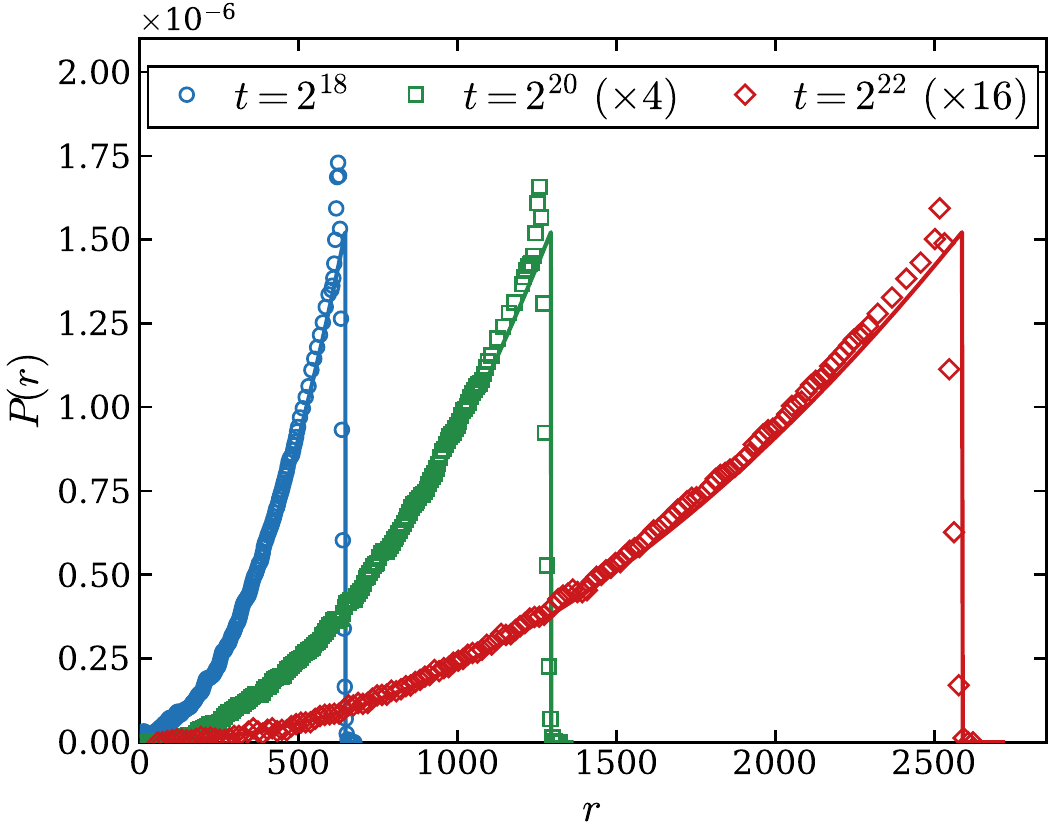}
    \caption{2D instantaneous probability density $P(r,t)$
    vs $r$ for $\alpha = 0.5 < \alpha_c = 1$ (interaction dominated regime). Analytical solutions Eq.~\eqref{eq:P_2d}
    (solid lines) agree with simulations (symbols) at three times
    $t = 2^{18}$, $2^{20}$, and $2^{22}$. Data for $t = 2^{18}$, $2^{20}$ are rescaled by
    $t/2^{18}$ to improve visibility. The characteristic $Z(y) \sim y^2$
    growth from the origin and sharp cutoff at $y_{\max}$ are
    clearly visible, confirming the annular particle distribution
    predicted by the theory. Deviations near $y_{\max}$ can be attributed to
    finite-time and finite-size corrections.}
    \label{fig:2D_without_diff}
\end{figure}

Figure~\ref{fig:2D_without_diff} shows $P(r)$ for $\alpha = 0.5 <
\alpha_c = 1$ at three times.
The agreement between theory and simulation is very good. The $Z(y) \sim y^2$ growth
from the origin and the sharp cutoff at $y_{\max}$ are clearly
reproduced by Eq.~\eqref{eq:P_2d}. Small deviations near the cutoff $y_{\max}$ can be attributed to finite-time and finite-size effects; they are expected to vanish as $t \to \infty$, $N \to \infty$.
We have also computed the integrated density $P_{\mathrm{int}}(r)$ for the same parameters. It agrees very well with the analytical result (\ref{eq:Pint_2d}) and qualitatively behaves analogously to the one-dimensional case shown in Fig.\ \ref{fig:1D_integ}.

\subsubsection{Marginal case ($\alpha = \alpha_c = 1$)}

In the marginal case, $\alpha = \alpha_c = 1$, the diffusion
makes a contribution to the motion for $t \to \infty$. Thus,  the full Fokker-Planck
equation~\eqref{eq:fokker_planck} must be retained.
Using $Z = -yY'/2$ and inserting the scaling forms, one obtains
a Riccati equation for $Z(y)$,
\begin{equation}
-\frac{1}{2}y^2 Z = -2A Z^2 + D y Z',
\label{eq:riccati_2d}
\end{equation}
which admits the closed-form solution
\begin{equation}
Z(y) = \frac{D \exp(-y^2/4D)}{D \cdot \mathcal{C}
+ A \cdot E_1(y^2/4D)},
\label{eq:Z_2d_riccati}
\end{equation}
where $E_1(x) = \int_x^\infty (e^{-u}/u)\,du$ is the
exponential integral, which has the asymptotic behavior
$E_1(z) \sim -\ln z - \gamma_E$ as $z \to 0^+$, with
$\gamma_E \approx 0.5772$ the Euler--Mascheroni constant. The constant
$\mathcal{C}$ is fixed by the normalization condition
$2\pi\int_0^\infty y Z(y)\,dy = 1$.
In terms of original coordinates,
\begin{equation}
P(r,t) = t^{-1}\,Z\!\left(\frac{r}{t^{1/2}}\right),
\end{equation}
which interpolates between logarithmic growth at small $r$ and
Gaussian decay $P \sim e^{-r^2/4Dt}$ at large $r$, without a sharp cutoff.

\begin{figure}
    \centering
    \includegraphics[width=\linewidth]{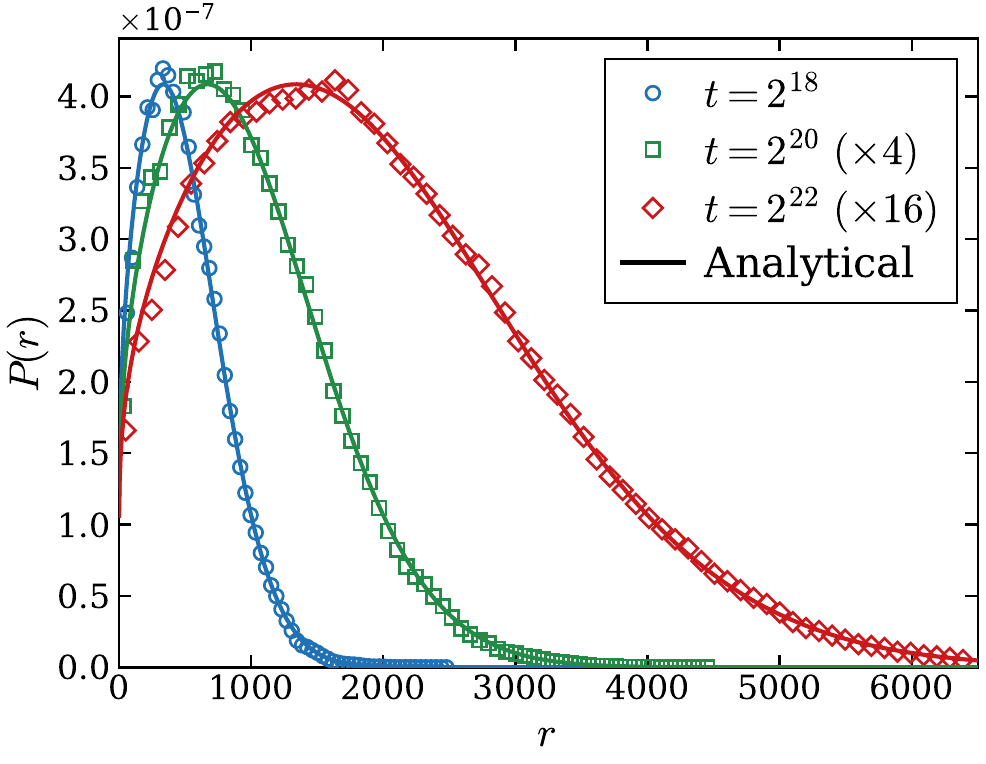}
    \caption{2D instantaneous probability density $P(r)$
    vs $r$ for the marginal case, $\alpha = \alpha_c = 1$.
    The analytical solution
    Eq.~\eqref{eq:Z_2d_riccati} (solid line) agrees with
    simulations (symbols) at three times $t = 2^{18}$, $2^{20}$,
    $2^{22}$, rescaled by $t/2^{18}$ for visual clarity.
    The Gaussian decay for large $r$ contrasts sharply with
    the hard cutoff seen in Fig.~\ref{fig:2D_without_diff}
    for $\alpha < \alpha_c$.}
    \label{fig:2D_with_diff}
\end{figure}

Figure~\ref{fig:2D_with_diff} demonstrates excellent quantitative
agreement between Eq.~\eqref{eq:Z_2d_riccati} and simulations at
$\alpha = \alpha_c = 1$ for three times. The characteristic Gaussian tail for large
$r$ is clearly visible, contrasting with the sharp cutoff seen
in Fig.~\ref{fig:2D_without_diff} for $\alpha < \alpha_c$.

\subsection{Three Dimensions}

With $d=3$, $\delta = 2/5$, $\alpha_c = 4/5$, the consistency
relation~\eqref{eq:consistency} gives $Z(y) = -Y/5 - 2yY'/5$,
and the normalization condition is
$4\pi \int_0^\infty y^2 Y(y)\,dy = 1$. Inserting the scaling
forms into the Fokker-Planck-like equation~\eqref{eq:fokker_planck_2} and applying the
consistency relation yields the nonlinear second-order ODE
\begin{equation}
\begin{split}
6Y + 18yY' + 4y^2Y'' = -A\bigg[&\frac{10Y'Y}{y} +
35(Y')^2 \\
&+ 5Y''Y + 20yY''Y'\bigg],
\end{split}
\label{eq:ODE_3d}
\end{equation}
A physical solution of Eq.~\eqref{eq:ODE_3d} must
fulfill the condition
$P_{\mathrm{int}}(r,t) = t^{-1/5}\, Y(r/t^{2/5}) \geq 0$
and the condition that $P_{\mathrm{int}}$ cannot decrease
with time. This can only be fulfilled if $Y(y)$ has a
divergence at $y = 0$.

The resulting $Y(y)$ has three distinct regimes.
Close to the origin ($y < y_0$), $Y(y)$ diverges as
$Y \sim y^{-1/2}$. In the intermediate regime
$y_0 < y \leq y_{\max}$, $Y(y)$ takes the quadratic
form $Y(y) = a_0 + a_2 y^2$. Finally, $Y(y)$ vanishes
for $y > y_{\max}$. The resulting scaling function of the integrated
density is
\begin{equation}
Y(y) = \begin{cases}
\dfrac{4y_{\max}^{5/2}}{25A \cdot 5^{1/4}}\, y^{-1/2}
& 0 < y < y_0\\[10pt]
\dfrac{y_{\max}^2 - y^2}{5A}
& y_0 \leq y \leq y_{\max}\\[10pt]
0 & y > y_{\max},
\end{cases}
\label{eq:Y_3d}
\end{equation}
where
$y_{\max} = [8 \pi(1+\sqrt{5}/125)/ (75 A)]^{-1/5}$ and $y_0 = y_{\max}/\sqrt{5}$.
Expressed in terms of $r$ and $t$, the integrated density reads
\begin{equation}
P_{\mathrm{int}}(r,t) = t^{-1/5}\, Y\!\left(\frac{r}{t^{2/5}}\right)~.
\label{eq:Pint_3d}
\end{equation}

The scaling function of the instantaneous density follows
from the consistency relation,
\begin{equation}
Z(y) = \begin{cases}
0 & y < y_0\\[6pt]
\dfrac{5y^2 - y_{\max}^2}{25A}
& y_0 \leq y \leq y_{\max}\\[10pt]
0 & y > y_{\max}.
\end{cases}
\label{eq:Z_3d}
\end{equation}
In terms of original coordinates,
\begin{equation}
P(r,t) = t^{-6/5}\, Z\!\left(\frac{r}{t^{2/5}}\right),
\label{eq:P_3d}
\end{equation}
which vanishes identically for $r < y_0\,t^{2/5}$, rises
quadratically for $y_0\,t^{2/5} < r < y_{\max}\,t^{2/5}$, and
cuts off sharply at the outer boundary $r = y_{\max}\,t^{2/5}$.
This gives rise to a \emph{spherical shell structure}: a depleted
core of radius $r_0 = y_0\,t^{2/5}$ surrounded by a shell
of particles near the outer boundary — a different
shape than the 1D bimodal or 2D annular cases, arising from
the increased geometric weight of large-$r$ regions in three
dimensions.

\begin{figure}
    \centering
    \includegraphics[width=\linewidth]{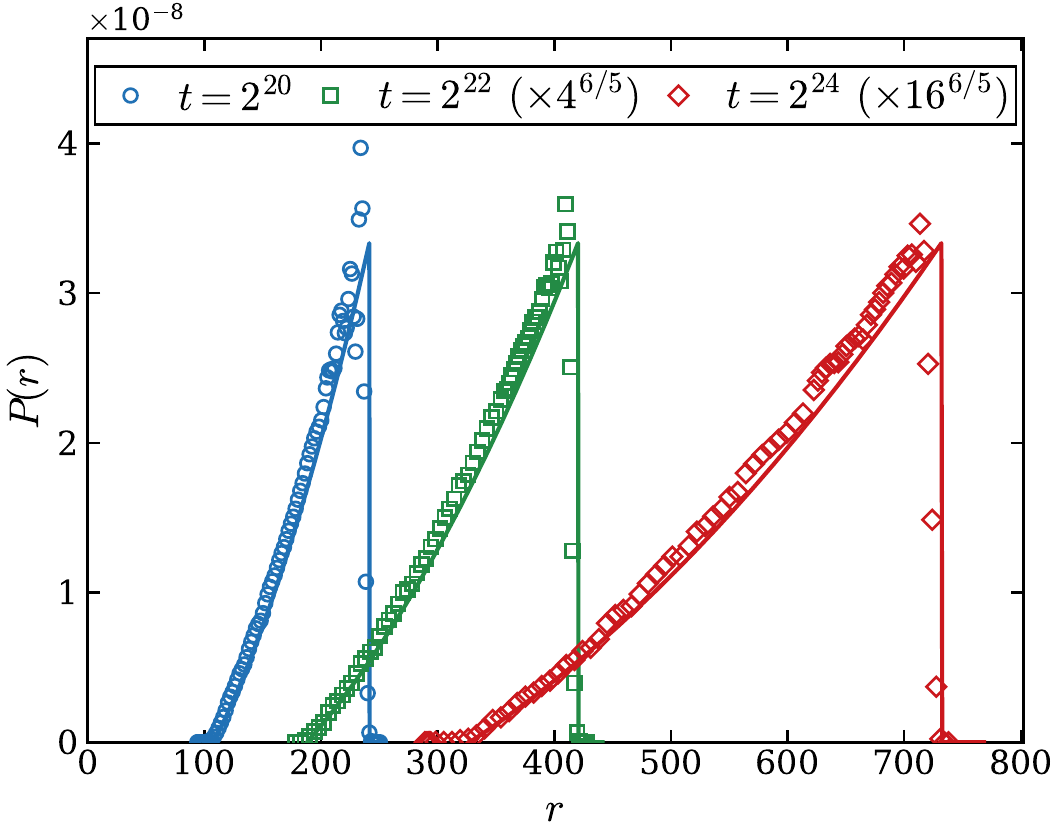}
    \caption{3D instantaneous probability density $P(r,t)$
    vs $r$ for $\alpha = 0.2 < \alpha_c = 4/5$. Analytical
    predictions Eq.~\eqref{eq:P_3d} (solid lines)
    match radial-binning simulations (symbols) at three times
    $t = 2^{18}$, $2^{20}$, $2^{22}$. The data are rescaled by
    $(t/2^{18})^{6/5}$ for better visibility. The spherical shell structure is clearly
    visible: a depleted core for $r < y_0\,t^{2/5}$ followed by a
    rising density terminating sharply at $r = y_{\max}\,t^{2/5}$.
    Deviations near $y_{\max}$ are finite-time and finite-ensemble corrections
    vanishing small $t \to \infty$, $N \to \infty$.}
    \label{fig:3D}
\end{figure}

\begin{figure}
    \centering
    \includegraphics[width=\linewidth]{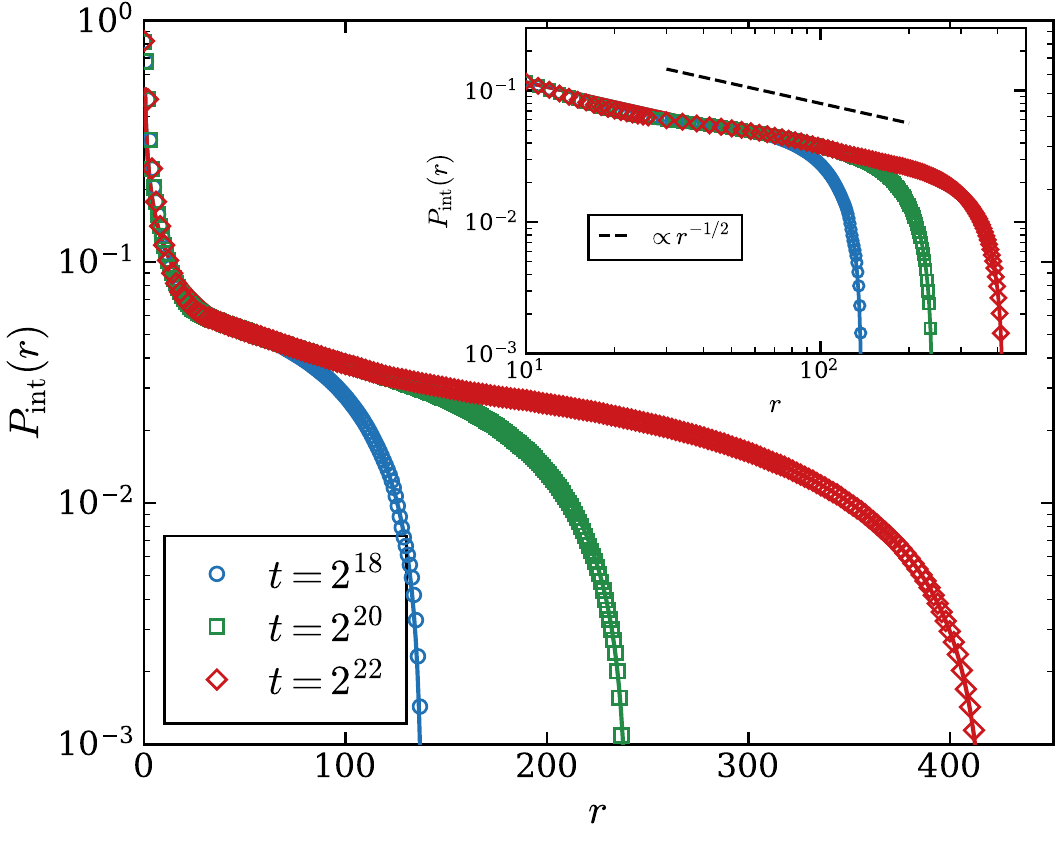}
    \caption{3D integrated probability density $P_{\mathrm{int}}(r,t)$
    vs $r$ for $\alpha = 0.2 < \alpha_c = 4/5$. Solid lines show the analytical
    predictions \eqref{eq:Pint_3d}. Inset: Double-logarithmic plot of the integrated density,
    demonstrating that $P_{\mathrm{int}}$ follows the predicted $r^{-1/2}$ power law
    (illustrated by the dashed line,  with arbitrary prefactor) for smaller $r$.
    }
    \label{fig:Pint_3D}
\end{figure}

Figure~\ref{fig:3D} shows $P(r)$ for $\alpha = 0.2 < \alpha_c = 4/5$
at three times $t = 2^{18}$, $2^{20}$, and $2^{22}$. The shell structure is evident in all three
curves, and the analytical predictions
\eqref{eq:P_3d} describe the data well.
As before, small deviations near $y_{\max}$ can be attributed to
finite-time and finite-ensemble-size effects.
The integrated density for the same parameters is shown in Fig.\ \ref{fig:Pint_3D}.
The agreement with the analytical predictions \eqref{eq:Pint_3d} is again very good. In particular, the data confirm the
power-law singularity regime for $y<y_0$. Note that the data very close to the origin correspond to early times and are therefore expected to feature corrections to the asymptotic scaling results arising from the analytical calculations. Such coreections are likely responsible for the small deviations between the data and Eq.\ \eqref{eq:Pint_3d} at small $r$.

\section{Conclusions}
\label{sec:conclusions}

In this paper we have developed an analytical mean-field theory
of myopic self-avoiding fractional Brownian motion,
a stochastic process in which an ensemble of particles undergoes
FBM while being repelled by the gradient of their own
time-integrated density~\cite{House2025}. Building on the
phenomenological scaling theory of Ref.~\cite{House2025}, we
derived closed-form solutions for the probability density in
one, two, and three dimensions in the interaction-dominated regime, and validated them against
large-scale numerical simulations.

To enable large-scale simulations in two and three dimensions,
we developed a radial binning algorithm that exploits the
rotational symmetry of the stationary state, reducing the
computational memory requirements from $\mathcal{O}(L^d)$
to $\mathcal{O}(L)$. This made three-dimensional simulations
of the process feasible for the first time. The algorithm was
validated by direct comparison with Cartesian binning in two
dimensions. Simulations for times up to $2^{22}$ confirmed
the analytical predictions for the probability density in all studied
dimensions.

The results of this work advance the modeling of axon growth
and self-organization in vertebrate brains. The power of these
stochastic models has been demonstrated in the research of
serotonergic fibers~\cite{Janusonis2019, Janusonis2020,
Janusonis2023, Janusonis2025}, a core component of biological
neural tissue, but this approach may also be applicable to
other strongly-stochastic axons classically known as the
``ascending reticular activating system''
(ARAS)~\cite{Korczyn2026, Brown2012}. This work also
contributes to the understanding of another large class of
systems that are characterized by stigmergy, a mechanism
whereby agents communicate across different time points by
leaving traces in their environment~\cite{Theraulaz1999}.
This behavior is often observed in biological systems as
diverse as immune cells, insect colonies, and human groups.
Importantly, this research has applications in the design
and targeted disruption of robotic-agent
swarms~\cite{Horvath2025}. In all of these systems, noise
and agent-to-agent interactions cooperate to achieve high
adaptability and robust self-organization in changing
environments. Analytical solutions democratize the use of
mathematical models in these research fields by obviating
the need to access high-performance computing resources.

Several directions remain open for future work.
Especially in the context of animal motion, it will
be of interest to study attractive mean-density
interactions, for instance, for animals leaving a
scent that attracts other individuals~\cite{Vilk2022}.
This would model collective behavior driven by
positive feedback, in contrast to the repulsive
self-avoidance studied here. Instead of the description in terms of the MSD, FBM can
also be analyzed in terms of the statistical properties
of single trajectories~\cite{Krapf2019}. It will be
interesting to apply this approach to myopic self-avoiding
FBM.

\par The
present mean-field theory assumes rotational symmetry and
treats the process in free space. A natural and biologically
relevant extension is to combine the mean-density interaction
with a reflecting boundary, which confines the process to a
finite region as is the case for serotonergic axons growing
within brain tissue. Preliminary simulations suggest that the
interplay of the reflecting wall and the mean-density interaction
qualitatively changes the process: the interaction cuts off the
power-law singularity observed in the probability density of
reflected FBM near the wall~\cite{Wada2018, Vojta2020}, and for
$\alpha > \alpha_c$ a fundamental breakdown of single-parameter
scaling emerges, creating two distinct dynamical regimes
separated by a time-dependent crossover length scale. This extended model, myopic self-avoiding FBM with a
reflecting boundary, is currently under investigation
and is expected to provide a more realistic framework
for the modeling of spatially bounded systems, including
serotonergic fibers constrained by the outer (pial) and
inner (ventricular) boundaries of brain tissue~\cite{Janusonis2019, Janusonis2020, Janusonis2023}.

\begin{acknowledgments}
This work was supported by the National Science Foundation
under Grant No. \#2112862. R.M.\ acknowledges support by
the German Science Foundation under Grant No.\ 318763901
CRC 1294 Data Assimilation, project B10. The simulations
were performed on the Pegasus and Mill clusters at
Missouri University of Science and Technology~\cite{MillHPC}.
\end{acknowledgments}

\appendix

\section{Cartesian vs.\ Radial Binning}
\label{sec:appendix}

The simulations require us to keep track of the $d$-dimensional
time-integrated density $P_{\mathrm{int}}(\mathbf{r}, t)$ at
every time step. In a straightforward Cartesian grid implementation,
this requires storing a full $d$-dimensional array whose size scales as
$L^d$, where $L$ is the linear system size. In one dimension,
binning along $x$ is straightforward and memory requirements
are modest. In two dimensions, binning in the $(x, y)$ plane
requires storing an $L \times L$ array, which is
memory-intensive (storage proportinonal $L^2$) and computationally
demanding. In three dimensions, a full Cartesian grid
(storage proportional to $L^3$ ) becomes prohibitively expensive and
in practice impossible for the system sizes needed to reach
the scaling regime.

To overcome this limitation, we exploit the rotational
symmetry of the density  of myopic self-avoiding
FBM. As established in Sec.~\ref{sec:model}, when all
particles start at the origin, $P_{\mathrm{int}}$ depends
only on the radial coordinate $r = |\mathbf{r}|$ and not
on the direction of $\mathbf{r}$. We therefore develop an
improved radial binning method: instead of storing the full
$d$-dimensional density array, we bin the density as a
function of $r$ only, reducing the memory and computational
complexity from $\mathcal{O}(L^d)$ to $\mathcal{O}(L)$.
Each particle's contribution is broadened via a two-dimensional or three-dimensional Gaussian of
width $\sigma_w = 0.5$ and then binned radially.
The radial force component $f_r(r,t) = -A\,\partial P_{\mathrm{int}}/
\partial r$ is computed from the resulting one-dimensional
radial density profile using fourth-order finite differences. This reduction in computer memory requirements makes large-scale three-dimensional simulations
in three dimensions feasible for the first time.

\begin{figure}[h]
    \centering
    \includegraphics[width=\linewidth]{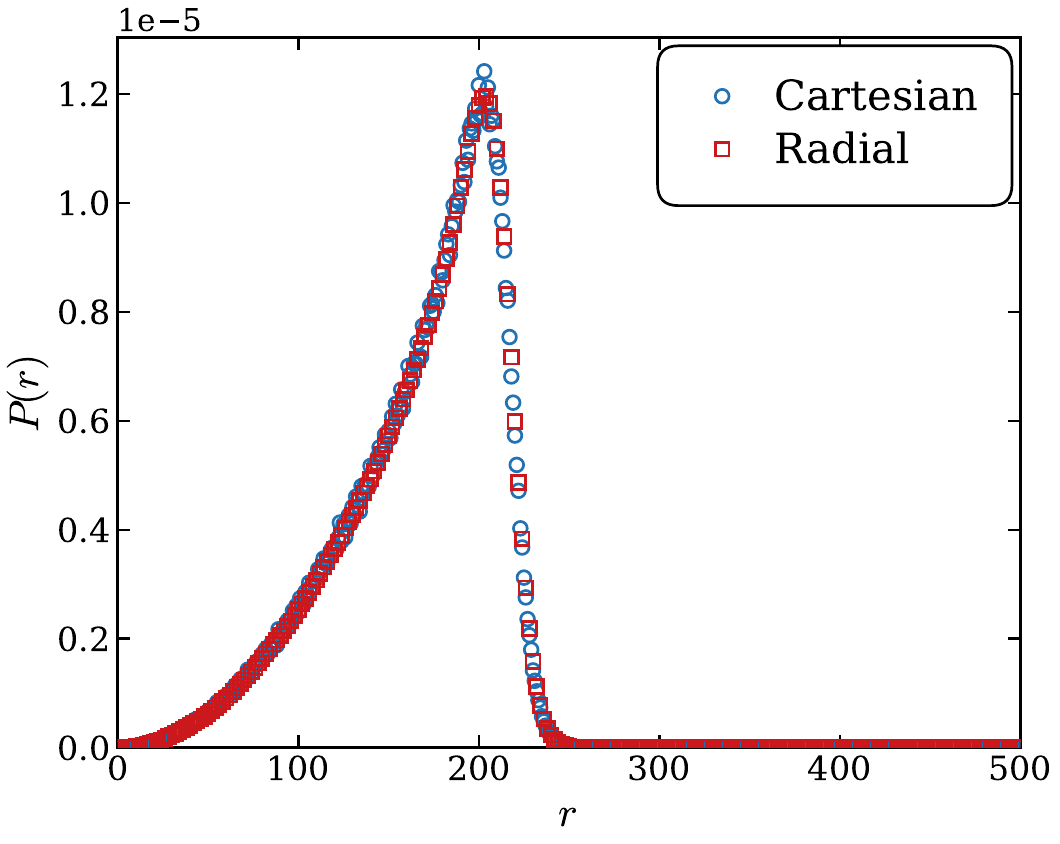}
    \caption{Comparison of 2D probability density  $P(r)$ vs $r$,
computed using Cartesian binning (blue circles) and radial
binning (orange squares) for $\alpha = 0.7 < \alpha_c = 1$,
at $t = 2^{18}$. The excellent agreement between both
methods validates the radial binning approach, which
reduces memory requirements from $\mathcal{O}(L^2)$
to $\mathcal{O}(L)$ and makes higher-dimensional
simulations feasible.}
\label{fig:2D_comparison}
\end{figure}

To validate this approach, we implemented both Cartesian and
radial binning in two dimensions with identical parameters,
where both methods are computationally feasible.
Figure~\ref{fig:2D_comparison} shows that both methods yield
indistinguishable density profiles across the full range of
$r$, confirming that the radial binning algorithm correctly
captures the density and forces without any loss of accuracy.

\newpage

\bibliography{references}
\end{document}